\documentclass[fleqn,usenatbib]{mnras}

\usepackage{newtxtext,newtxmath}
\usepackage{hyperref}
\usepackage[T1]{fontenc}
\usepackage{ae,aecompl}

\usepackage{graphicx}	% Including figure files
\usepackage{amsmath}	% Advanced maths commands
\usepackage{amssymb}	% Extra maths symbols

\newcommand\snameone{EPIC 219654213}
\newcommand\vsini{$v$\,sin\,$i_1$}   

\newcommand\teff{$T_{\rm{eff}}$}

\newcommand\Msun{\hbox{$M_{\odot}$}}  %Msun
\newcommand\Rsun{\hbox{$R_{\odot}$}}  %Rsun
\newcommand\kms{\hbox{km\,s$^{-1}$}}  %km per sec
\newcommand{\PERIOD}{$ 5.441995   \pm 0.000007$}
\newcommand{\EPOCH}{$ 2471.1452   \pm 0.0002$}
\newcommand{\ECCENTRICITY}{$0.0073 \pm 0.0003$}

\newcommand{\SSMA}{$9.07\pm 0.03$}
\newcommand{\SSR}{$ 0.1320 \pm 0.0003$}

\newcommand{\CONJUNCTION}{$ 0.05 \pm 0.04$}
\newcommand{\INCLINATION}{$ 89.70 \pm 0.23$}
\newcommand{\LDOne}{$ 0.6$ (fixed)}
\newcommand{\LDTwo}{$0.2$ (fixed)}

\newcommand{\BRIGHTNESSRATIO}{$0.026 \pm 0.005$}
\newcommand{\KAmp}{$20.066 \pm 0.009$}
\newcommand{\KAmpLC}{$13.7 \pm 0.7$}
\newcommand{\GAMMA}{$-17.57 \pm 0.07$}
\newcommand{\OFFa}{$0.18 \pm 0.07$}
\newcommand{\OFFb}{$-0.95 \pm 0.15$}
\newcommand{\OFFc}{$30.48 \pm 0.07$}
\newcommand{\PrimaryAlbedo}{$ 0.71 \pm 0.17 $}
\newcommand{\SecondaryAlbedo}{$ 0.30 \pm 0.20 $}

\newcommand{\PrimaryMass}{$1.008 \substack{+0.081 \\ -0.097}$}
\newcommand{\PrimaryRadius}{$1.518 \substack{+0.038 \\ -0.049}$}
\newcommand{\SecondaryMass}{$0.187 \substack{+0.012 \\ -0.013}$}
\newcommand{\SecondaryRadius}{$0.200 \substack{+0.007 \\ -0.008}$}
\newcommand{\EAONE}{$0.00059 \pm 0.00001$}
\newcommand{\EATWO}{$-0.00006 \pm 0.00004$}
\newcommand{\EATHREE}{$0.00005 \pm 0.00004$}
\newcommand{\EAFOUR}{$-0.00001 \pm 0.00004$}
\newcommand{\EAFIVE}{$0.00007 \pm 0.00004$}

\newcommand{\SMA}{$ 0.065\pm 0.003$}
\newcommand{\DURATION}{$ 4.58 \pm 0.02$}

\newcommand{\PrimaryAge}{$4.1 \pm 1.1$}

\newcommand{\PrimaryRadiusspecmatch}{$1.58 \pm 0.18$}
\newcommand{\PrimaryTeff}{$6305 \pm 110$}
\newcommand{\PrimaryMet}{$-0.08 \pm 0.09$}
\newcommand{\Primaryvsini}{$15.08 \pm 0.62$}
\newcommand{\PrimaryType}{F7V}
\newcommand{\SecondaryType}{M5V}

\title[A transiting M-dwarf showing beaming effect in the field of Ruprecht 147]{A transiting M-dwarf showing beaming effect in the field of Ruprecht 147}

\author[Ph. Eigm\"uller et. al]
{Philipp Eigm\"uller,$^{1,2}$\thanks{E-mail: philipp.eigmueller@dlr.de}
Szil\'ard Csizmadia,$^{1}$
Michael~Endl,$^{3}$
Davide Gandolfi,$^{4}$
\newauthor
William~D.~Cochran,$^{3}$
David Yong,$^{5}$
Alexis M. S. Smith,$^{1}$
Juan~Cabrera,$^{1}$
\newauthor 
Hans J. Deeg,$^{6,7}$
Marshall C. Johnson,$^{8}$
Judith Korth, $^{9}$
Jorge Prieto-Arranz,$^{6,7}$
\newauthor 
David Nespral, $^{6,7}$
Artie P. Hatzes, $^{10}$
\\
$^{1}$Institute of Planetary Research, German Aerospace Center, Rutherfordstrasse 2, 12489 Berlin, Germany\\
$^{2}$Center for Astronomy and Astrophysics, TU Berlin, Hardenbergstr. 36, D-10623 Berlin, Germany\\
$^{3}$Department of Astronomy and McDonald Observatory, University of Texas at Austin, 2515 Speedway, Stop C1400, Austin, TX 78712, USA\\
$^{4}$Dipartimento di Fisica, Universit\'a di Torino, via P. Giuria 1, 10125 Torino, Italy\\
$^{5}$Research School of Astronomy and Astrophysics, The Australian National University, Cotter Road, Canberra, ACT 2611, Australia\\
$^{6}$Instituto de Astrof\'\i sica de Canarias, C. V\'\i a L\'actea S/N, E-38205 La Laguna, Tenerife, Spain\\
$^{7}$Universidad de La Laguna, Dept. de Astrof\'\i sica, E-38206 La Laguna, Tenerife, Spain\\
$^{8}$Department of Astronomy, The Ohio State University, 140 West 18th Ave., Columbus, OH 43210, USA\\
$^{9}$Rheinisches  Institut  f\"ur  Umweltforschung  an  der  Universit\"at  zu K\"oln, Aachener Strasse 209, 50931 K\"oln, Germany\\
$^{10}$Th\"uringer Landessternwarte Tautenburg, D-07778 Tautenburg, Germany}
\date{Accepted XXX. Received YYY; in original form ZZZ}

\pubyear{2016}

\begin{document}
\label{firstpage}
\pagerange{\pageref{firstpage}--\pageref{lastpage}}
\maketitle

\begin{abstract}

We report the discovery and characterization of an eclipsing \SecondaryType~dwarf star, orbiting a slightly evolved \PrimaryType~main sequence star.

In contrast to previous claims in the literature, we confirm that the system does not belong to the galactic open cluster Ruprecht 147.
We determine its fundamental parameters combining K2 time-series data with spectroscopic observations from the McDonald Observatory, FIES@NOT, and HIRES@KECK. The very precise photometric data from the K2 mission allows us to measure variations caused by the beaming effect (relativistic doppler boosting), ellipsoidal variation, reflection, and the secondary eclipse. We determined the radial velocity using spectroscopic observations and compare it to the radial velocity determined from the beaming effect observed in the photometric data. The \SecondaryType~star has a radius of \SecondaryRadius~\Rsun~and a mass of \SecondaryMass~\Msun. The primary star has radius of \PrimaryRadius~\Rsun~and a mass of \PrimaryMass~\Msun. The orbital period is \PERIOD~days. The system is one of the few eclipsing systems with observed beaming effect and spectroscopic radial velocity measurements and it can be used as test case for the modelling of the beaming effect. 

Current and forthcoming space missions such as TESS and PLATO might benefit of the analysis of the beaming effect to estimate the mass of transiting companions without the need for radial velocity follow up observations, provided that the systematic sources of noise affecting this method are well understood. 
\end{abstract}

%%%%%%%%%%%%%%%%%%%%%%%%%%%%%%%%%%%%%%%%%%%%%%%%%%

%%%%%%%%%%%%%%%%% BODY OF PAPER %%%%%%%%%%%%%%%%%%

\section{Introduction}

To understand the evolution of stars and planetary systems it is fundamental to derive observationally the fundamental parameters of stars in different stages of their evolution and compare those results with stellar evolution models. 
Although low mass stars with a mass well below one solar mass are most common in our solar neighborhood, they are not yet completely understood, even in regards to their bulk parameters.
They show significant discrepancies between theoretical and observed mass radius relation. For very low mass stars (VLMSs) in a mass regime between $0.1~M_{\odot}$ and $0.6~M_{\odot}$ \citet{Mann2015} found that Dartmouth models \citep[][]{Dotter2008} systematically underestimate the radius by $\approx~4.7\%$ and overestimate the effective temperature by $\approx~2.2\%$. \\
One key observational method to determine the mass and radius of low mass stars is the study of detached eclipsing binaries (DEBs). The DEBCat catalog \citep[][]{Southworth2015} of DEBs lists currently 29\footnote{http://www.astro.keele.ac.uk/jkt/debcat; state of June 2018} well characterized  VLMSs with a mass below $0.6~M_{\odot}$. DEBcat is limited to DEBs with their bulk parameters determined to a precision better than $2\%$. Many more DEBs not as well characterized are known \citep[e.g.][]{Eigmuller2016, Gillen2017, Chaturvedi2018}.

In this paper we present the detailed characterization of a DEB formed by a main sequence star and a M dwarf companion with precise \textit{K2} photometry and ground-based radial velocity follow-up.

We deduce the bulk characteristics of an M dwarf companion to a solar like star modeling \textit{K2} light curve and radial velocity follow up measurements. The high precision light curves by the Kepler satellite allow us not only to model the primary eclipse but also to model the occultation as well as reflection, ellipsoidal variation, and the relativistic beaming effect. Due to the high contrast ratio between late and early type stars, secondary eclipse of M dwarfs are only rarely observed in such systems. The observation of the secondary eclipse and reflection allows us to give additional constrains on the luminosity ratio in the binary system, and the albedo of the M-dwarf. The ellipsoidal variation depends mainly on the mass ratio of the two components and the semi major axis of the system, and thus also gives further constrains on the system parameters.

\subsection{Relativistic Beaming}
The relativistic beaming effect is caused by the reflex motion of the stars introducing photometric flux variations due to the Doppler effect. The theoretical background of the relativistic beaming effect has been discussed for eclipsing binary stars \citep{Zucker2007} as well as for planetary systems \citep{Loeb2003}. Using light curves of CoRoT and the Kepler satellite a few observations of this effect have been reported in the last years \citep[][]{Mazeh2010, Kerkwijk2010, Bloemen2011, Herrero2014, Faigler2015, TalOr2015}. For a few transiting systems \citep{Mazeh2010,Bloemen2011, Faigler2015} and even more non-transiting systems \citep{TalOr2015} spectroscopic radial velocity measurements are available.\\

The measurement of the relativistic beaming effect allows an independent estimate of the radial velocity of the secondary component of the binary system, which can be used to  establish the nature of the companion and to determine the mass ratio between primary and secondary object.

This effect has been proposed in the literature as a tool to confirm the nature of transiting planetary companions, which otherwise typically require an extensive ground-based follow-up campaign to confirm their nature. 
The scheduling of ground-based resources is one of the current challenges for space-borne transit surveys like \textit{K2} or, in the future missions, TESS and PLATO. 
Understanding the limitations of relativistic beaming effect will allow to establish this method as an independent tool to identify low mass stellar companions, one of the main sources of false-positives for transit surveys.
Unfortunately, as it will be shown in this paper, the current state-of-the-art approach neglects the influence of stellar variability, which might compromise the retrieval of the radial velocity value from the photometry \citep{Faigler2015, Csizmadia2018}. In case of disagreement between the radial velocity amplitudes between photometry and spectroscopy, the latter value is preferred.

\section{Observations}
\subsection{K2 photometry and transit detection}

The Kepler space observatory, launched in 2009, was designed to provide precise photometric monitoring of over $150,000$ stars in a single field and to detect transiting Earth-sized planets with orbital periods up to one year \citep{2010Sci...327..977B}. Due to failure of two reaction wheels the Kepler mission stopped after four years of operation. At the end of 2013 the operation of the Kepler space telescope re-started with a new concept that uses the remaining reaction wheels, the spacecraft thrusters, and solar wind pressure, to point the telescope. The new mission, called K2 \citep{K2}, enables the continued use of the Kepler spacecraft with limited pointing accuracy. K2 observes different fields located along the ecliptic for a duration of about three consecutive months per field. Ruprecht 147 is an open cluster observed with K2 during Campaign 7. It was observed using a super-aperture, tiled with 60 51x51 masks, totaling 156,060 pixels.\\

For the photometry we combined the single masks to the super-aperture for all 4043 frames. A master frame, combined out of all 4043 frames, was used to identify sources and their individual masks using the label function in the python scipy module. The photometry was performed using a fixed aperture for each object as described in \citet{Eigmuller2017}. Similar to the Kepler pipeline of  \citet{Vanderburg2014}, each light curve was split in segments to remove noise correlated with the pointing of the Kepler spacecraft. For the transit detection we used the DST algorithm described in \citet{cabrera2012}. This algorithm has been largely used by our team to detect planets in other K2 fields \citep{Barragan2016,Grziwa16,Johnson2016,Smith2017,Eigmuller2017}. \snameone \mbox{} was identified as possible planetary candidate and radial velocity follow up observations were scheduled to determine the mass of the companion. These observations allowed us to characterize the companion as \SecondaryType \mbox{} star.\\

\begin{figure}
	\centering
	\includegraphics[width=0.5\columnwidth]{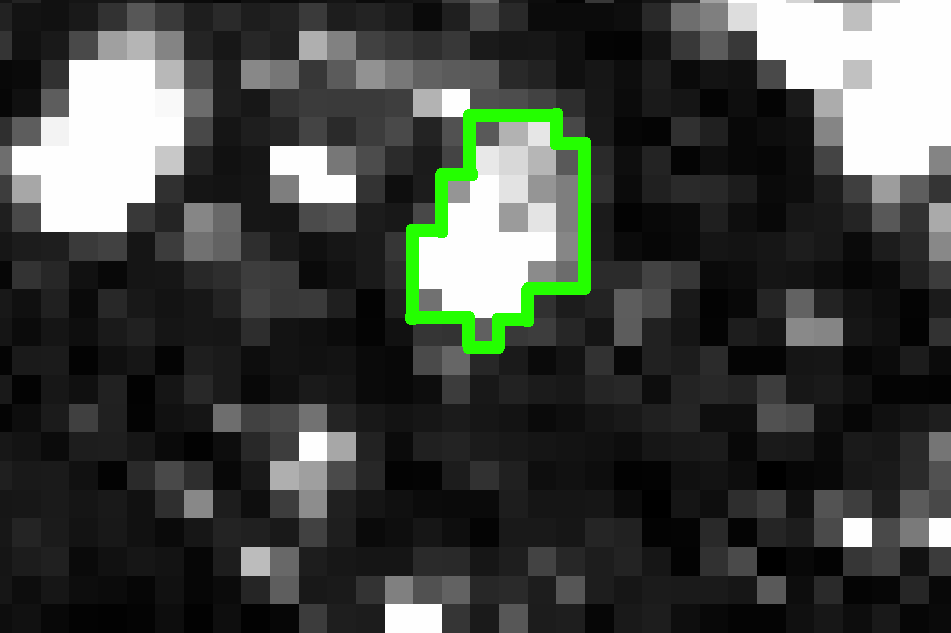}
    \caption{Stamp of K2 observation of this object. The green line shows the aperture selected for photometry.}
    \label{fig::fov}
\end{figure}

\begin{figure}
	\includegraphics[width=\columnwidth]{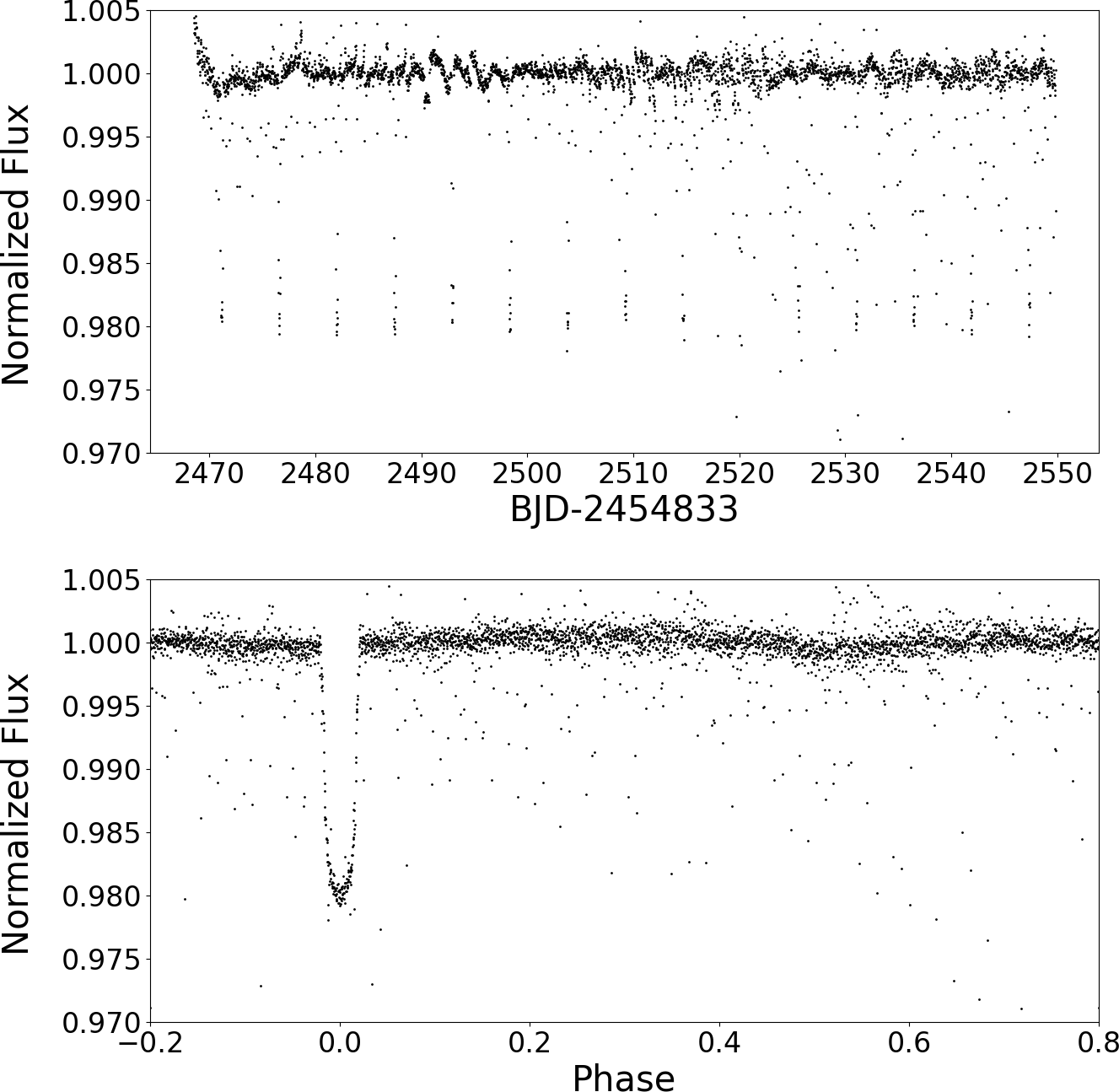}
    \caption{Corrected and normalized light curve of \snameone. The upper plot shows the raw normalized light curve over time, with longterm trends removed. The lower plot displays the same data, phase folded to the orbital period.}
    \label{fig:lightcurve}
\end{figure}

For the modelling of the light curve we re-analyzed the photometry with optimal selected aperture size and   customized segment size for de-correlation. We tested different aperture sizes and finally selected by manual inspection the aperture as shown in Figure~\ref{fig::fov}. The size of the segments for decorrelation and detrending have been selected to be twice the orbital period of the EB. This way we avoid splitting the light curve within any eclipse signal. These short segments were individually de-correlated against the relative motion of the star, given in the \texttt{POS\_CORR} columns. To remove long term trends we de-correlated these segments also in the time domain for linear trends. The resulting light curve, in the time domain and phase folded, is shown in Figure~\ref{fig:lightcurve}. In a last step we removed outliers and the ramp at the beginning of the light curve.\\

\subsection{High Dispersion Spectroscopy}
\label{RV_Follow_up}
Spectroscopic radial velocity follow up has been carried out to verify the nature of the companion. We used the Tull Spectrograph at the McDonald Observatory, the FIES spectrograph at the Nordic Optical Telescope and the HIRES Spectrograph at the Keck Observatory.

\subsection{Tull spectrograph @ McDonald Observatory}
We observed EPIC~219654213 with the Tull Coud\'e spectrograph \citep{Tull1995} at the Harlan J. Smith 2.7 m telescope at McDonald Observatory. The Tull spectrograph images the entire optical spectrum on its CCD detector at a resolving power of $R=60,000$. Between 2016 July and 2017 September we collected a total of 8 spectra of the star. Exposure times ranged from 20 minutes to 1 hour, depending on seeing conditions. The data were flat-fielded, bias-subtracted and wavelength-calibrated using standard IRAF procedures. The extracted spectra have a S/N-ratio at 5650~\AA\, between 19 and 36 per 2-pixel resolution element. \\
For each observation we computed an absolute radial velocity of the star by cross-correlating it with the RV standard star HD~182572 with an absolute RV of -100.35\,km\,s$^{-1}$ \citep{Jofre2015}. The RV data are listed in Table \ref{rvs} and displayed in Figure \ref{fig:rvfitphase}.

\subsection{FIES @ Nordic Optical Telescope}
We observed \snameone\ with the the FIbre-fed \'Echelle Spectrograph \citep[FIES;][]{Telting2014} mounted at the 2.56m Nordic Optical Telescope (NOT) of Roque de los Muchachos Observatory (La Palma, Spain). As part of the observing programs 54-205, 54-027, and 55-019, we secured 7 intermediate-resolution ($R$\,$\approx$\,47\,000) spectra between October 4, 2016, and May 22, 2017 UTC. The exposure time was set to 1200-3600 sec, based on our scheduling constrains and sky conditions. We followed the same observing strategy described in \citet{Gandolfi2015} and traced the RV drift of the instrument by bracketing the science exposures with long-exposed ($T_\mathrm{exp}$\,=\,35-60 sec) ThAr spectra. We reduced the data using standard IRAF and IDL routines and extracted the RVs via multi-order cross-correlations with the co-added spectrum of the star. \\
The extracted spectra have a S/N ratio between 10 and 20 per pixel at 5500\,\AA.

\subsection{HIRES @ Keck Observatory}
HIRES \citep{Vogt1994} observations were obtained on 1 and 2 June 2016 UTC using the blue
configuration. We used the B5 slit (width = 0.86", length = 3.5") which
provided a resolving power of R=49,000, and the CCD binning was 2x2. The
echelle and cross-disperser angles were set to 0.0 and 0.97, respectively; this
provided wavelength coverage from ~3030\AA\, to ~5900\AA. The observing 
sequence was \snameone (600s), ThAr lamp, then the radial velocity
standard HD 182572 ($2 \times 5s$). Data reduction was performed using
MAKEE\footnote{MAKEE was developed by T. A. Barlow specifically for reduction
of Keck HIRES data. It is freely available at 
\url{http://www2.keck.hawaii.edu/realpublic/inst/hires/data_reduction.html}}. 

An additional HIRES observation was obtained on 14 Aug 2016 UTC using the red
configuration. The B2 slit (width = 0.57", length = 7.0") was used providing a
resolution of R=66,000, and the CCD binning was 1x1. The echelle and
cross-disperser angles were set to 0.0 and 0.45, respectively, resulting in a
wavelength coverage from ~4200\AA\, to ~8500\AA. The observing sequence was \snameone (600s), the radial velocity standard HD 182572 ($2 \times 5s$), then a
ThAr lamp. Data reduction was performed using MAKEE. For all exposures, the S/N
was ~40-50 per pixel near 5500\AA\, for \snameone. For both sets of
observations, the RV was measured by multi-order cross-correlation against HD
182572 \citep[assuming $-$100.35 \kms;][]{Udry1999} using the FXCOR package in
IRAF \citep{Tody1993}.

\begin{table}
\caption{Tull, FIES and HARPS-N RV measurements of \snameone.\label{rvs}}
\begin{tabular}{lccr}
\hline
\hline
BJD$_\mathrm{TDB}$ & RV & $\sigma_{\mathrm{RV}}$ &   Instr. \\
$-$2,450,000 & (\kms) & (\kms)  &  \\
\hline
\noalign{\smallskip}
7541.06966767     & -13.13   & 0.16	  & HIRES Blue \\
7542.03177372     &  1.67    & 0.65	  & HIRES Blue \\
7542.85072551     &  -2.87   & 0.68   & Tull \\
7614.8006962      &  -28.5   &  0.6   & HIRES Red \\
7666.39011879     & 20.604   & 0.040  & FIES \\
7668.37494413     &  21.605 & 0.027   & FIES \\
7669.37951759     & -0.053  &  0.071  & FIES \\
7672.60931308     &  1.79    & 1.19   & Tull \\
7673.60167106     &  -4.45   & 1.09   & Tull \\
7893.67142075     & -4.773  & 0.094   & FIES \\
7894.69572934     &  14.638 & 0.074   & FIES \\
7895.64675401     & 31.723  & 0.043   & FIES \\
7896.70668445     & 26.071  & 0.067   & FIES \\
7954.80808902     &   -9.93  & 0.46   & Tull \\
7994.70153193     &   -4.84  & 0.89   & Tull \\
8008.62752647     &  -23.38  & 0.76   & Tull \\
8009.62419535     &   -3.09  & 0.78   & Tull \\
8010.70251742     &   -0.24  & 1.77   & Tull \\
\hline
\end{tabular}
\end{table}

\section{Analysis}
\subsection{Spectral Analysis}
The primary star has been characterized using the SpecMatch-emp tool by \citet{Yee2017} on a combined FIES spectrum. We empirically determined the effective temperature, metallicity and radius of the host star. The result is in agreement with an independent spectral analysis using the KEA tool \citep{Endl2016} of a single spectrum taken with the Tull spectrograph. The main parameters of the primary star are listed in Table \ref{tab:stellar}. The effective temperature is \PrimaryTeff \, K, and the radius is \PrimaryRadiusspecmatch \,\Rsun.

\begin{table*}[!th]
\begin{center}
\caption{Main identifiers, coordinates, magnitudes, and spectroscopic parameters of the Primary Star of \snameone.\label{tab:stellar}}
\begin{tabular}{lcc}
\hline
\hline
\noalign{\smallskip}
Parameter & {\texttt{\snameone}} & Unit\\
\noalign{\smallskip}
\hline
\noalign{\smallskip}
RA 	& 19$^h$17$^m$00$^s$.670   & h \\
DEC & -16$\degr$11$^\prime$32$^{\prime \prime}$.53  & deg\\
UCAC4 ID & 370-168815  & \ldots\\
EPIC ID & 219654213  & \ldots\\
pm RA (GAIA DR2) & $1.209 \pm 0.048$ 	&  $mas\,yr^{-1}$\\
pm DEC (GAIA DR2) & $-8.619 \pm 0.042$ & $mas\,yr^{-1}$\\
parallax (GAIA DR2) & $0.7930 \pm 0.0331$	& $mas$\\

\noalign{\smallskip}
\hline
\noalign{\smallskip}
Effective Temperature \teff & \PrimaryTeff  & K\\
Stellar Radius $R_1$ & \PrimaryRadiusspecmatch & Rsun\\
Metallicity [Fe/H] & \PrimaryMet    & dex \\
\vsini\footnote{The \vsini we retrieved from spectral modeling using the KEA spectral modelling tool} & \Primaryvsini &  \kms \\
Age         			& \PrimaryAge                 &Gyr\\
Spectral Type & G2\,V & \ldots\\
\noalign{\smallskip}
\hline
\noalign{\smallskip}
B mag (UCAC4) & $14.533 \pm 0.05$ & mag\\
V mag (UCAC4) & $13.911 \pm 0.08$ & mag\\
J mag (2MASS) & $12.697 \pm 0.03$ & mag\\
H mag (2MASS) & $12.467 \pm 0.02$ & mag\\
K mag (2MASS) & $12.426 \pm 0.02$ & mag\\
\noalign{\smallskip}
\hline
\end{tabular}
\end{center}
\end{table*}
\subsection{Joint Analysis of Photometric and Radial Velocity Measurements}
We used the Transit Light Curve Modelling (\texttt{TLCM}) code\footnote{We used the latest version which was tlcm92\_f (June, 2018).}  \citep{Csizmadia2015, Csizmadia2018} for the analysis of the light curves and radial velocity measurements. \texttt{TLCM} is based on the \cite{M&A} model to fit planetary transit light curves. The RV measurements are modeled with a Keplerian orbit. The fit is optimized using first a genetic algorithm and then a simulated annealing chain. Uncertainties were estimated from MCMC chains. \texttt{TLCM} is capable of modeling the out of transit variations caused by ellipsoidal variation, reflection and relativistic beaming. To model the beaming the index factors $\alpha$  were calculated by convolving ATLAS spectra at the temperature of the star with the sensitivity curve of K2. Then the flux variations are given as $(1+\frac{2}{3}\alpha \cdot V_{rad}/c)$  for every phase. The factor of $\frac{2}{3}$ comes from the spherical geometry. \\

The fitted parameters are the scaled
semi-major axis $a/R_*$ and 
radius ratio  $R_\mathrm{p}/R_*$, 
the conjunction parameter $i$, 
the eccentricity given by $e\cdot sin(\omega)$ and $e\cdot cos(\omega)$, 
the limb darkening coefficients $u_+=u_1+u_2$ and  $u_- = u_1 -u_2$, 
the brightness ratio, 
the radial velocity semi amplitude $K$ (from spectroscopic radial velocity data and the photometric beaming effect combined, but clearly dominated by the spectroscopic data) and 
the systemic $\gamma$-velocity, 
the ellipsoidal variability, represented by five terms of Legendre polynoms with amplitudes $a_j$, and
the albedo of both components. 
The period ($P_\mathrm{orb}$) and 
epoch of mid-transit ($T_0$) are allowed to vary slightly around the values determined already by the detection. 
The contamination factor was determined using the GAIA DR2 information of background objects within our aperture.
We also fitted for radial velocity trends that might unveil the presence of additional orbiting companions in the systems. We obtained radial accelerations that are consistent with zero. 

\begin{figure}
	\includegraphics[width=\columnwidth]{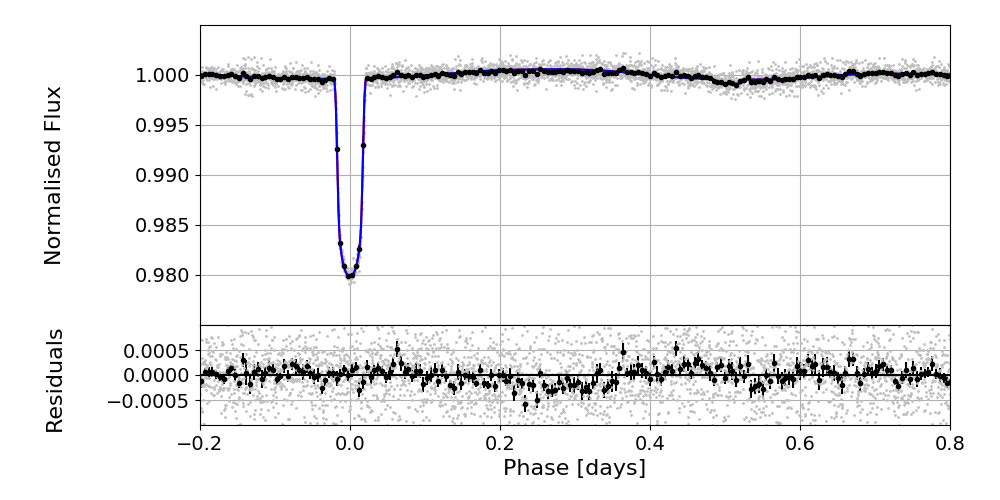}
    \caption{Phase folded light curve and the best fitting transit model of \snameone. Gray points are the measurements, black circles the binned data.  The continuous  line represents the best fitted model. Residuals to the fit are shown in the lower panel.}
    \label{fig:lcfit}
\end{figure}

\begin{figure}
	\includegraphics[width=\columnwidth]{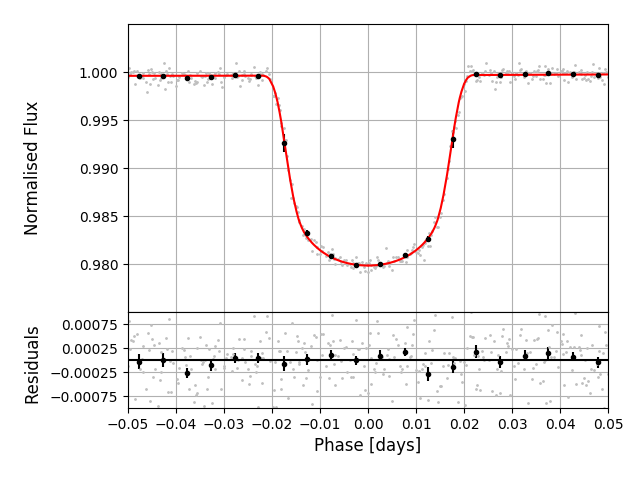}
    \caption{Phase folded light curve of the primary transit and the best fitting transit model of \snameone. Gray points are the measurements, black circles the binned data. The continuous red line represents the best fitted model. Residuals to the fit are shown in the lower panel.}
    \label{fig:lcfitzoom}
\end{figure}

\begin{figure}
	\includegraphics[width=\columnwidth]{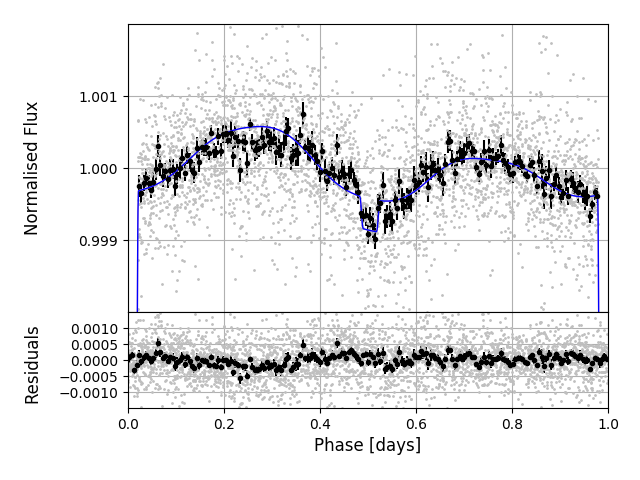}
    \caption{Phase folded light curve with the y-scale adapted to highlight the secondary eclipse and out-of-transit variations. Gray points are the measurements, black circles the binned data. The continuous blue line represents the best fitted model. Residuals to the fit are shown in the lower panel.}
    \label{fig:lcfitzoom2}
\end{figure}

\begin{figure}
	\includegraphics[width=\columnwidth]{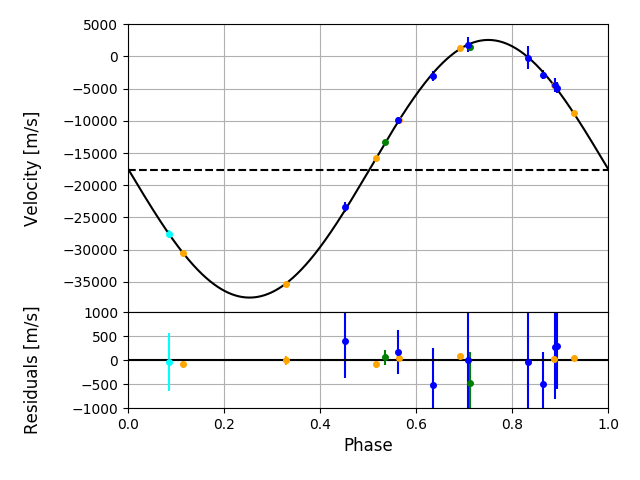}
    \caption{Phase folded RV measurements of \snameone\, (Tull: blue; HIRES Blue: green; HIRES Red: cyan; FIES: yellow) and best fitting model (black line). Residuals to the fit are shown in the lower panel.}
    \label{fig:rvfitphase}
\end{figure}

The best fitting transit model is shown in Figure~\ref{fig:lcfit}, along with the photometric data. Figures~\ref{fig:lcfitzoom} and \ref{fig:lcfitzoom2} show expanded views around the primary  and secondary transit, respectively. The RV data are shown in Figure~\ref{fig:rvfitphase}.
Results of the combined modelling  are listed in Table \ref{tab:param}. 

To retrieve the mass of the primary star we compared Dartmouth stellar models \citep{Dotter2008}\footnote{Characterization of the primary star using Parsec1.2S  isochrones \citep{Bressan2012} gave results in agreement with our findings using the Dartmouth stellar models.}.
With the effective temperature, metallicity, and radius from spectral classification using SpecMatch-emp, we select the according Dartmouth stellar models consistent with the stellar density according to our modelling results.
\\

\begin{table*}[!th]
\begin{center}
\caption{Parameters from light curve and RV data analysis. Mass and radius of both components determined by combining modelling results with Dartmouth stellar models.\label{tab:param}}
\begin{tabular}{lcc}
\hline
\hline
\noalign{\smallskip}
Parameter & Modelling Result&  Unit\\
\noalign{\smallskip}
\hline
\noalign{\smallskip}
Orbital period $P_\mathrm{orb}$  & 		\PERIOD	 & days\\
Transit epoch $T_0$ 			 & 		\EPOCH	 & BJD$_\mathrm{TDB}-2450000$\\
Transit Duration 				 & 		\DURATION	 & hours\\
Scaled semi-major axis $a/R_*$ 	 &    	\SSMA	& \\
Semi-major axis $a$ 			 &    \SMA	 &au\\
Scaled Secondary radius $R_2/R_1$ 	 &  \SSR  & \\
Orbital inclination angle $i$	 &    \INCLINATION    &$\deg$\\
Conjunction parameter $b_c$ 			 &    \CONJUNCTION     & \\
Limb-darkening coefficient $u_+$ &   \LDOne  &\\
Limb-darkening coefficient $u_-$ &   \LDTwo  &\\
Brightness ratio & \BRIGHTNESSRATIO  &\\
Albedo of primary star & \PrimaryAlbedo &\\
Albedo of secondary star & \SecondaryAlbedo &\\
Ellipsoidal variability term $a_1$ & \EAONE &\\
Ellipsoidal variability term $a_2$ & \EATWO &\\
Ellipsoidal variability term $a_3$ & \EATHREE &\\
Ellipsoidal variability term $a_4$ & \EAFOUR &\\
Ellipsoidal variability term $a_5$ & \EAFIVE &\\
\noalign{\smallskip}
\hline
\noalign{\smallskip}
Radial velocity semi amplitude $K$ &  \KAmp  &$ \kms $\\
Systemic radial velocity $\gamma$ & \GAMMA & \kms  \\
RV velocity offset FIES / Tull &  \OFFc     &\kms \\
RV velocity offset HIRES Blue / Tull & \OFFa   &  \kms \\
RV velocity offset HIRES Red / Tull &  \OFFb   &\kms \\
Eccentricity $e$ & \ECCENTRICITY                &\\
\noalign{\smallskip}
\hline
%\hline
%\noalign{\smallskip}
%\multicolumn{3}{c}{Bulk parameters of both components}\\
\noalign{\smallskip}
%\hline
Primary mass $M_1$     			& \PrimaryMass   &\Msun\\
Primary radius $R_1$   			& \PrimaryRadius &\Rsun\\
%$M^{1/3}_1/R_1$  					&       \MoverR         & &Solar units\\
%Primary mean density $\rho_1$ & \PrimaryRho    			 & &\gcm \\
%Primary surface gravity \logg\footnote{\label{note}Derived from the light curve modeling, effective temperature, metal content, and isochrones.} & \PrimaryLogg   & &cgs\\
%Distance               & & & parsec\\
\noalign{\smallskip}
Secondary mass $M_2$   			& \SecondaryMass &\Msun\\
Secondary radius $R_2$ 			& \SecondaryRadius &\Rsun\\
%Secondary mean density          & & &\gcm\\
%Secondary surface gravity \loggp         			& \SecondaryLogg  &cgs\\
%Secondary calculated effective temperature & \SecondaryTeff &K\\
\noalign{\smallskip}
\hline
\end{tabular}
\end{center}
\end{table*}

\section{Discussion}
The binary system \snameone\ consists of a \PrimaryType\mbox{ } main sequence star orbited by an \SecondaryType\mbox{ } star on nearly circular orbit with an orbital period of \PERIOD\mbox{ } days. We can confirm a small but significant eccentricity of \ECCENTRICITY. The radial velocity semi amplitude of K=\KAmp\mbox{ }\kms together with the mass of the primary star (\PrimaryMass\Msun) leads to a mass of the M dwarf companion of \SecondaryMass\Msun. The radius of the M dwarf is \SecondaryRadius\Rsun. The uncertainties in stellar parameters are model dependent as the bulk parameters of the primary star are based on Dartmouth isochrones.\\ 

\subsection{Ruprecht\,147 cluster membership}
The open cluster Ruprecht\,147 is an old nearby clusters \citep{Curtis2013}. Open clusters are a laboratory for stellar astrophysics. For members of open clusters we can constrain their age which would allow to give much better constrains for stellar evolution theories. This is why it is important to know whether \snameone\, belongs to Ruprecht\,147. The cluster membership for Ruprecht\,147 has  been analyzed in the past as part of global catalogs of open clusters \citep[e.g.][]{Kharchenko2005,Dias2006,Dias2014} and in detail by \citet{Curtis2013}. There is a discrepancy between results by  \citet{Dias2014}, who gives \snameone \mbox{ } a cluster membership probability of 83\%, and   \citet{Curtis2013}, who does not list \snameone \mbox{ } as a cluster member. \\
Using the GAIA DR2 catalogue \citep{GAIADR2} we confirm the result of \citet{Curtis2013}. Based on the GAIA DR2 five-parameter astrometry we identify 102 Cluster members which allow us to determine the distance of the Ruprecht 147 cluster to $310 \pm 20$ pc. The mean proper motion of the cluster is $pm_{ra} = -0.84 \pm 0.84\, mas\, yr^{-1}$, $pm_{dec}= -26.7 \pm 0.9\, mas\, yr^{-1}$. The GAIA-DR2 astrometric solution for \snameone \mbox{ } however gives a distance of 
 $1260 \pm 50$ pc and a proper motion of $pm_{ra} = 1.21 \pm 0.05\, mas\, yr^{-1}$, and $pm_{dec} = -8.62 \pm 0.04\, mas\, yr^{-1}$ which is in agreement with \snameone \mbox{ } being a field star.

\subsection{Relativistic beaming effect}\label{sec::beaming}
As we can clearly see ellipsoidal variation, reflection, and the beaming effect in our light curve we also modeled the light curve data without taking the spectroscopic RV data into account to compare the radial velocity determined spectroscopically to the radial velocity determined from the beaming effect. In this case we fixed the eccentricity to the value determined by our combined modelling as the depth of the secondary eclipse is close to the noise level and does not allow us to restrict the eccentricity 
\footnote{We also modelled the light curve with the eccentricity fixed to $e=0$ as a test case. As the eccentricity of the system is very small it had no major impact on the final results. To allow better comparison with our combined modelling we thus fixed the eccentricity to the value determined from the radial velocity follow up observations.}. The beaming factor is estimated using relations to the effective temperature and metallicity of the star \citep[cf.][]{Csizmadia2018}.\\

Fitting the light curve independent of the RV data, the best fitted radial velocity semi-amplitude is $K_{LC}= $\KAmpLC \kms\, in comparison to the spectroscopic radial velocity $K =$ \KAmp \kms\, when fitting spectroscopic and photometric data together. The results are not in agreement with each other. The difference in the amplitude corresponding to the beaming effect of the respective radial velocity amplitude is $\Delta A_{Beam}\approx 75 ppm$.\\
In our combined model the radial velocity semi-amplitude is dominated by the spectroscopic data, allowing us to use the combined model to determine the radial velocity semi amplitude without being impaired by this discrepancy. \\

A possible explanation for the difference in radial velocity measured by the beaming effect and spectroscopic data might be additional variability in the light curve which is not included in our model. Our light curve model includes  ellipsoidal variation, reflection, and the relativistic beaming effect. Additional brightness variations on the surface of the primary star might introduce further rotational variability, which in case of synchronized rotation would have the same period as the other effects. The rotational velocity of the primary star, estimated from \vsini\, and stellar radius, is in agreement with the system being tidally locked.  As our model does not include rotational variability, the presence of such signal would be compensated by adapting the ellipsoidal variation, the beaming effect and the reflection.\\

As we expect rotational variability to be changing over timescales of a few rotations we tested this assumption by splitting the light curve into two segments and modelling these segments independently. The results show a change in the amplitude of the beaming effect of 35ppm, which account for 50\% of the observed discrepancy between expected and observed beaming effect. This shows that the systematic error in modelling the beaming effect is much larger than is reflected in the value of $K_{LC}= $\KAmpLC \kms, but still might not explain the whole offset. For rotational variability to be causing this offset there would need to be a contribution that is stable over the whole observation, which might be if it is introduced by  interactions between the two stars. \\

We tested whether the uncertainty in third light diluting our signal could also explain the observed discrepancy, but the effect would be an order of magnitude to small.

\subsection{Low Mass Stars Mass-Radius relation}  
Using spectroscopic radial velocity measurements in combination with a high precision light curve we were able to determine the mass and radius of the companion with only a few percent uncertainty. The mass radius relation of the M-dwarf is in the range predicted by stellar models for stars of this age. \snameone B\, with a mass of \SecondaryMass\Msun\, is one of only few characterized very low mass stars in eclipsing binaries.   

\section{Conclusions}

We discovered an eclipsing binary star \snameone,  consisting of an \SecondaryType \, orbiting an \PrimaryType\, star.  \snameone\, is not a member of the cluster Ruprecht\,147.
High precision photometry by the K2 mission and radial velocity data allowed to characterize the system and both components. The \SecondaryType \, star has a mass of $M_2=$\,\,\SecondaryMass \Msun\, and a radius of $R_2=$ \SecondaryRadius \Rsun. The primary star has an mass of $M_1=$ \PrimaryMass \Msun \, and a radius of $R_2=$\PrimaryRadius \Rsun. The high precision photometry allowed us to observe also the photometric beaming effect. Its amplitude is not in agreement with the radial velocity measured spectroscopically. 
Using the beaming effect to determine the mass of the secondary objects gives $M_{2LC} = 0.13 \pm 0.04$ \Msun\,, which underestimates the mass by $\approx$35\%. 
However, detailed analysis of the light curve showed that the amplitude of the out of transit variation  changes with time which might hint towards the presence of additional variability in our light curve, preventing us from using beaming to estimate the mass of the companion.
Especially for short period binaries, where the rotational period is synchronized with the orbital period this might be a common effect.
This shows how careful one has to treat the beaming effect when using it to determine the mass of the secondary object. However, for surveys spanning over several orbital periods, a detailed analysis might help to estimate the mass, taking additional stellar variability into account.\\

The upcoming TESS and future PLATO mission are expected to deliver large numbers of planetary  candidates, thus resources for spectroscopic follow up will be limited. For cases such as \snameone\, mass estimate of the secondary object will be needed to distinguish between the companion being an highly inflated hot Jupiter, late M-dwarf, and brown dwarf. It therefore has been proposed to classify such systems without spectroscopic radial velocity follow up by using the beaming effect. 
Our analysis shows that this needs to be done with care. Additional variability needs to be taken into account and might prevent us from proper modelling of the beaming effect.

\section*{Acknowledgements}
This paper includes data collected by the K2 mission. Funding for the K2 mission is provided by the NASA Science Mission directorate.\\

This paper is based on observations made with the Nordic Optical Telescope, operated by the Nordic Optical Telescope Scientific Association at the Observatorio del Roque de los Muchachos, La Palma, Spain, of the Instituto de Astrofisica de Canarias.\\

This paper includes data taken at The McDonald Observatory of The University of Texas at Austin.\\

Some of the data presented herein were obtained at the W. M. Keck Observatory, which is operated as a scientific partnership among the California Institute of Technology, the University of California and the National Aeronautics and Space Administration. The Observatory was made possible by the generous financial support of the W. M. Keck Foundation. The authors wish to recognize and acknowledge the very significant cultural role and reverence that the summit of Maunakea has always had within the indigenous Hawaiian community.  We are most fortunate to have the opportunity to conduct observations from this mountain. \\

This work has made use of data from the European Space Agency (ESA)
mission {\it Gaia} (\url{https://www.cosmos.esa.int/gaia}), processed by
the {\it Gaia} Data Processing and Analysis Consortium (DPAC,
\url{https://www.cosmos.esa.int/web/gaia/dpac/consortium}). Funding
for the DPAC has been provided by national institutions, in particular
the institutions participating in the {\it Gaia} Multilateral Agreement. \\

HJD acknowledges support by grant ESP2015-65712-C5-4-R of the Spanish Secretary of State for R\&D\&i (MINECO).\\

M.E. and W.D.C. were supported by NASA grant NNX16AE70G to The University of Texas at Austin.\\

%%%%%%%%%%%%%%%%%%%%%%%%%%%%%%%%%%%%%%%%%%%%%%%%%%



%\bibliographystyle{mnras}

\begin{thebibliography}{99}
%\bibitem[\protect\citeauthoryear{Aerts et al.}{2010}]{Aerts2010} Aerts, C., Christensen-Dalsgaard, J., \& Kurtz, D.~W.\ 2010, Asteroseismology, Astronomy and Astrophysics Library.~ISBN 978-1-4020-5178-4.~Springer Science+Business Media B.V., 2010, p.,  

%\bibitem[\protect\citeauthoryear{Andersen}{1991}]{Andersen1991} Andersen, J.\ 1991, \aapr, 3, 91 
\bibitem[\protect\citeauthoryear{Barrag{\'a}n et al.}{2016}]{Barragan2016} Barrag{\'a}n, O., Grziwa, S., Gandolfi, D., et al.\ 2016, \aj, 152, 193 
\bibitem[\protect\citeauthoryear{Borucki et al.}{2010}]{2010Sci...327..977B} Borucki, W.~J., Koch, D., Basri, G., et al.\ 2010, Science, 327, 977 
\bibitem[\protect\citeauthoryear{Bloemen et al.}{2011}]{Bloemen2011} Bloemen, S., Marsh, T.~R., {\O}stensen, R.~H., et al.\ 2011, \mnras, 410, 1787 
\bibitem[\protect\citeauthoryear{Bressan etal.}{2012}]{Bressan2012}{Bressan}, A., {Marigo}, P., {Girardi}, L., {et~al.} 2012, \mnras, 427, 127
\bibitem[\protect\citeauthoryear{Cabrera et al.}{2012}]{cabrera2012} Cabrera, J., Csizmadia, Sz., Erikson, A., Rauer, H., \& Kirste, S.\ 2012, \aap, 548, A44 
%\bibitem[\protect\citeauthoryear{Carter \& Winn}{2009}]{carter2009} Carter, J.~A., \& Winn, J.~N.\ 2009, \apj, 704, 51 

\bibitem[\protect\citeauthoryear{Chaturvedi et al.}{2018}]{Chaturvedi2018} Chaturvedi, P., Sharma, R., Chakraborty, A., Anandarao, B.~G., \& Prasad, N.~J.~S.~S.~V 2018, arXiv:1805.05841 
%\bibitem[\protect\citeauthoryear{Cochran et al.}{2015}]{Cochran2015} Cochran, W.~D., Endl, M., Johnson, M.~C., et al.\ 2015, AAS/Division for Planetary Sciences Meeting Abstracts, 47, 417.02 
\bibitem[\protect\citeauthoryear{Csizmadia et al.}{2015}]{Csizmadia2015} Csizmadia, Sz., Hatzes, A., Gandolfi, D. et al. 2015, \aap, 584, A13
\bibitem[\protect\citeauthoryear{Csizmadia}{submitted}]{Csizmadia2018} Csizmadia, Sz., submitted to  \mnras

\bibitem[\protect\citeauthoryear{Curtis et al.}{2013}]{Curtis2013} Curtis, J.~L., Wolfgang, A., Wright, J.~T., Brewer, J.~M., \& Johnson, J.~A.\ 2013, \aj, 145, 134 
%\bibitem[\protect\citeauthoryear{Curtis}{2016}]{Curtis2016} Curtis, J.~L.\ 2016, Ph.D.~Thesis,  
%\bibitem[\protect\citeauthoryear{Delorme et al.}{2011}]{Delorme2011} Delorme, P., Collier Cameron, A., Hebb, L., et al.\ 2011, \mnras, 413, 2218 
\bibitem[\protect\citeauthoryear{Dias et al.}{2006}]{Dias2006} Dias, W.~S., Assafin, M., Fl{\'o}rio, V., Alessi, B.~S., \& L{\'{\i}}bero, V.\ 2006, \aap, 446, 949 
\bibitem[\protect\citeauthoryear{Dias et al.}{2014}]{Dias2014} Dias, W.~S., Monteiro, H., Caetano, T.~C., et al.\ 2014, \aap, 564, A79 
\bibitem[\protect\citeauthoryear{Dotter et al.}{2008}]{Dotter2008} Dotter, A., Chaboyer, B., Jevremovi{\'c}, D., et al.\ 2008, \apjs, 178, 89-101 
\bibitem[\protect\citeauthoryear{Eigm{\"u}ller et al.}{2016}]{Eigmuller2016} Eigm{\"u}ller, P., Eisl{\"o}ffel, J., Csizmadia, S., et al.\ 2016, \aj, 151, 84 
\bibitem[\protect\citeauthoryear{Eigm{\"u}ller et al.}{2017}]{Eigmuller2017} Eigm{\"u}ller, P., Gandolfi, D., Persson, C.~M., et al.\ 2017, \aj, 153, 130 
\bibitem[\protect\citeauthoryear{Endl \& Cochran}{2016}]{Endl2016} Endl, M., \& Cochran, W.~D.\ 2016, \pasp, 128, 094502 
\bibitem[\protect\citeauthoryear{Faigler et al.}{2015}]{Faigler2015} Faigler, S., Kull, I., Mazeh, T., et al.\ 2015, \apj, 815, 26 
%\bibitem[\protect\citeauthoryear{Fridlund et al.}{2017}]{Fridlund2017} Fridlund, M., Gaidos, E., Barrag{\'a}n, O., et al.\ 2017, \aap, 604, A16 
\bibitem[\protect\citeauthoryear{Gaia Collaboration et al.}{2016}]{GAIA2016} Gaia Collaboration, Brown, A.~G.~A., Vallenari, A., et al.\ 2016, \aap, 595, A2 
\bibitem[\protect\citeauthoryear{Gandolfi et al.}{2015}]{Gandolfi2015} Gandolfi, D., Parviainen, H., Deeg, H.~J., et al.\ 2015, \aap, 576, A11 
%\bibitem[\protect\citeauthoryear{Gandolfi et al.}{2017}]{Gandolfi2017} Gandolfi, D., Barrag{\'a}n, O., Hatzes, A.~P., et al.\ 2017, \aj, 154, 123 
\bibitem[\protect\citeauthoryear{Gaia Collaboration et al.}{2018}]{GAIADR2} Gaia Collaboration, Brown, A.~G.~A., Vallenari, A., et al.\ 2018, arXiv:1804.09365 
\bibitem[\protect\citeauthoryear{Gillen et al.}{2017}]{Gillen2017} Gillen, E., Hillenbrand, L.~A., David, T.~J., et al.\ 2017,  
%\bibitem[\protect\citeauthoryear{Girardi et al.}{2000}]{Girardi2000} Girardi, L., Bressan, A., Bertelli, G., \& Chiosi, C.\ 2000, \aaps, 141, 371 
\bibitem[\protect\citeauthoryear{Grziwa et al.}{2016a}]{Grziwa16} Grziwa, S., Gandolfi, D., Csizmadia, Sz., et al.\ 2016a, \aj, 152, 132
\bibitem[\protect\citeauthoryear{Herrero et al.}{2014}]{Herrero2014} Herrero, E., Lanza, A.~F., Ribas, I., et al.\ 2014, \aap, 563, A104 
\bibitem[\protect\citeauthoryear{Howell et al.}{2014}]{K2} Howell, S.~B., Sobeck, C., Haas, M. et~al., 2014, \pasp, 126, 398
%\bibitem[\protect\citeauthoryear{Jackson et al.}{2008}]{Jackson2008} Jackson, B., Greenberg, R., \& Barnes, R.\ 2008, \apj, 678, 1396-1406 
\bibitem[\protect\citeauthoryear{Jofr{\'e} et al.}{2015}]{Jofre2015} Jofr{\'e}, E., Petrucci, R., Saffe, C., et al.\ 2015, \aap, 574, A50 
\bibitem[\protect\citeauthoryear{Johnson et al.}{2016}]{Johnson2016} Johnson, M.~C., Gandolfi, D., Fridlund, M., et al.\ 2016, \aj, 151, 171 
\bibitem[\protect\citeauthoryear{Kharchenko et al.}{2005}]{Kharchenko2005} Kharchenko, N.~V., Piskunov, A.~E., R{\"o}ser, S., Schilbach, E., \& Scholz, R.-D.\ 2005, \aap, 438, 1163 

\bibitem[\protect\citeauthoryear{van Kerkwijk et al.}{2010}]{Kerkwijk2010} van Kerkwijk, M.~H., Rappaport, S.~A., Breton, R.~P., et al.\ 2010, \apj, 715, 51 
%\bibitem[\protect\citeauthoryear{Kiraga \& Stepien}{2007}]{Kiraga2007} Kiraga, M., \& Stepien, K.\ 2007, \actaa, 57, 149 

%\bibitem[\protect\citeauthoryear{Leconte et al.}{2010}]{Leconte2010} Leconte, J., Chabrier, G., Baraffe, I., \& Levrard, B.\ 2010, \aap, 516, A64 
\bibitem[\protect\citeauthoryear{Loeb \& Gaudi}{2003}]{Loeb2003} Loeb, A., \& Gaudi, B.~S.\ 2003, \apjl, 588, L117 
%\bibitem[\protect\citeauthoryear{Lucy \& Sweeney}{1971}]{Lucy1971} Lucy L.~B., Sweeney M.~A., 1971, AJ, 76, 544 
\bibitem[\protect\citeauthoryear{Mandel \& Agol}{2002}]{M&A} Mandel, K., and Agol, E. 2002, \apjl, 580, L171
\bibitem[\protect\citeauthoryear{Mann et al.}{2015}]{Mann2015} Mann, A.~W., Feiden, G.~A., Gaidos, E., Boyajian, T., \& von Braun, K.\ 2015, \apj, 804, 64 
%\bibitem[\protect\citeauthoryear{Marigo et al.}{2008}]{Marigo2008} Marigo, P., Girardi, L., Bressan, A., et al.\ 2008, \aap, 482, 883 
\bibitem[\protect\citeauthoryear{Mazeh \& Faigler}{2010}]{Mazeh2010} Mazeh, T., \& Faigler, S.\ 2010, \aap, 521, L59 
%\bibitem[\protect\citeauthoryear{R{\"o}ser et al.}{2008}]{Roeser2008} R{\"o}ser, S., Schilbach, E., Schwan, H., et al.\ 2008, \aap, 488, 401 
%\bibitem[\protect\citeauthoryear{Nowak et al.}{2017}]{Nowak2017} Nowak, G., Palle, E., Gandolfi, D., et al.\ 2017, \aj, 153, 131 
\bibitem[\protect\citeauthoryear{Smith et al.}{2017}]{Smith2017} Smith, A.~M.~S., Gandolfi, D., Barrag{\'a}n, O., et al.\ 2017, \mnras, 464, 2708 
%\bibitem[\protect\citeauthoryear{Smith et al.}{2017)}]{Smith2017} Smith, A.~M.~S., Cabrera, J., Csizmadia, S., et al.\ 2017, arXiv:1707.04549 
\bibitem[\protect\citeauthoryear{Southworth} {2015}]{Southworth2015} Southworth, J.\ 2015, Living Together: Planets, Host Stars and Binaries, 496, 164 
%\bibitem[\protect\citeauthoryear{Stelzer et al.}{2013}]{Stelzer2013} Stelzer, B., Marino, A., Micela, G., L{\'o}pez-Santiago, J., \& Liefke, C.\ 2013, \mnras, 431, 2063
\bibitem[\protect\citeauthoryear{Tal-Or et al.}{2015}]{TalOr2015} Tal-Or, L., Faigler, S., \& Mazeh, T.\ 2015, \aap, 580, A21 
\bibitem[\protect\citeauthoryear{Telting et al.}{2014}]{Telting2014} Telting, J.~H., Avila, G., Buchhave, L., et al.\ 2014, Astronomische Nachrichten, 335, 41
\bibitem[\protect\citeauthoryear{Tody}{1993}]{Tody1993} Tody, D.\ 1993, Astronomical Data Analysis Software and Systems II, 52, 173 
\bibitem[\protect\citeauthoryear{Tull et al.}{1995}]{Tull1995} Tull, R.~G., MacQueen, P.~J., Sneden, C., \& Lambert, D.~L.\ 1995, \pasp, 107, 251 
\bibitem[\protect\citeauthoryear{Udry et al.}{1999}]{Udry1999} Udry S., Mayor M., Queloz D., 1999, ASPC, 185, 367 
\bibitem[\protect\citeauthoryear{Vanderburg \& Johnson}{2014}]{Vanderburg2014} Vanderburg, A., \& Johnson, J.~A.\ 2014, \pasp, 126, 948 
\bibitem[\protect\citeauthoryear{Vogt et al.}{1994}]{Vogt1994} Vogt, S.~S., Allen, S.~L., Bigelow, B.~C., et al.\ 1994, \procspie, 2198, 362 
\bibitem[\protect\citeauthoryear{Yee et al.}{2017}]{Yee2017} Yee, S.~W., Petigura, E.~A., \& von Braun, K.\ 2017, \apj, 836, 77 
\bibitem[\protect\citeauthoryear{Zucker et al.}{2007}]{Zucker2007} Zucker, S., Mazeh, T., \& Alexander, T.\ 2007, \apj, 670, 1326 

\end{thebibliography}


%%%%%%%%%%%%%%%%%%%%%%%%%%%%%%%%%%%%%%%%%%%%%%%%%%

%%%%%%%%%%%%%%%%% APPENDICES %%%%%%%%%%%%%%%%%%%%%
%
%\appendix

% Don't change these lines
\bsp	% typesetting comment
\label{lastpage}
\end{document}